 \documentclass[preprint2]{aastex}

\shorttitle{Accretion disk spectrum of LMC X-3}
\shortauthors{Kubota et al.}

\begin{document}


\title{Testing Accretion Disk Structure with Suzaku data of LMC X-3}

\author{Aya Kubota\altaffilmark{1},  Chris Done\altaffilmark{2}, Shane W. Davis\altaffilmark{3}, Tadayasu Dotani\altaffilmark{4},
Tsunefumi Mizuno\altaffilmark{5} and Yoshihiro Ueda\altaffilmark{6}
}

\altaffiltext{1}{Department of Electronic Information Systems, Shibaura Institute of Technology, 307 Fukasaku, Minuma-ku, Saitama-shi, Saitam 337-8570, Japan}
\altaffiltext{2}{Department of Physics, University of Durham, South Road,
Durham, DH1 3LE, UK}
\altaffiltext{3}{School of Natural Sciences, Institute for Advanced Study, Einstein Drive, Princeton, NJ 08540, USA, Chandra Fellow}
\altaffiltext{4}{Institute of Space and Astronautical Science, 
Japan Aerospace Exploration Agency, 
3-1-1 Yoshinodai, Sagamihara, Kanagawa 229-8510, Japan
}
\altaffiltext{5}{Department of Physics, Hiroshima University,
1-3-1 Kagamiyama, Higashi-Hiroshima, Hiroshima 739-8526, Japan
}
\altaffiltext{6}{
Department of Astronomy, Kyoto University, Sakyo-ku, Kyoto 606-8502, Japan
}
\email{aya@shibaura-it.ac.jp}



\begin{abstract}

The Suzaku observation of LMC X-3 gives the best data to date on the
shape of the accretion disk spectrum. This is due to the combination
of very low absorbing column density along this line of sight which
allows the shape of the disk emisison to be constrained at low
energies by the CCD's, while the tail can be simultaneously determined
up to 30~keV by the high energy detectors. These data clearly
demonstrate that the observed disk spectrum is broader than a simple
`sum of blackbodies', and relativistic smearing of the emission is
strongly required. However, the intrinsic emission should be more
complex than a (color-corrected) sum of blackbodies as it should also
contain photo-electric absorption edges from the partially ionised
disk photosphere. These are broadened by the relativistic smearing,
but the models predict $\sim$ 3--5 per cent deviations for 1/3--1 solar abundance
around the edge
energies,  significantly stronger than observed. 
This indicate that the models need to include more physical processes such as
self-irradiation, bound-bound (line) absorption, and/or emission from
recombination continuua and/or lines. Alternatively, if none of these
match the data, it may instead require that the accretion disk density
and/or emissivity profile with height is different to that
assumed. Thus these data demonstrate the feasibility of observational
tests of our fundamental understanding of the vertical structure of
accretion disks.

\end{abstract}





\section{Introduction}

 It is important to study the emission from the accretion disk to understand how the gravitational energy is converted to radiation.
The simplest models of accretion disk spectra assume that the
gravitational energy dissipated at each radius thermalizes to a
blackbody spectrum. Summing it  over all radii, under an appropriate
inner boundary condition, produces the well--known Shakura
\& Sunyaev (1973) disk model. This is derived in Newtonian gravity,
but was extended to full general relativity by \cite{novikov73}. 
The maximum temperature produced by a disk accreting close to
the Eddington limit around a $10~M_\odot$ black hole is 1--2~keV,
easily observable with X-ray satellites, whereas a similarly Eddington
limited Active Galactic Nuclei (AGN) of $10^7M_\odot$ has a
temperature of 30--60~eV, in the unobservable EUV regime. Thus
stellar remnant black hole binary (hereafter BHB) systems give a much better
test of disk models than AGN.

The emission from each radius is a true blackbody only when the
disk is effectively optically thick to absorption at all frequencies.
Free-free (continuum) absorption drops as the frequency increases,  so
the highest energy photons from each radii are unlikely to
thermalize. This forms instead a modified (or diluted) blackbody, with
effective temperature higher than that for complete thermalization by a factor of $f_{\rm col}$ (termed a color
temperature correction). The
full disk spectrum is then a sum of these modified blackbodies, but
this can likewise be approximately described by a single color temperature
correction to a sum of blackbody disk spectrum\citep{st95}. 
However, continuum processes do not necessarily dominate the total absorption at
all frequencies. Bound-free (photo-electric) absorption from partially
ionised metals can be important, especially at high frequencies where
the free-free absorption becomes less significant \citep{dav05}. This imprints
atomic features onto the emission from each radius, distorting the
spectrum from a smooth continuum. The strength of these features is
set by radiative transfer through the vertical structure of the
photosphere, so they are one of the few diagnostics of the internal
disk properties \citep{dd08}.

Thus the intrinsic spectrum from each radius can be complex, but each
of these is smeared out by the combination of special and general
relativisic effects which arise from the rapid rotation of the
emitting material in a strong gravitational field \citep{cunningham75}.
The resultant smeared spectra are summed together to form the total
disk emission, which is not that different from a sum of smeared
color temperature corrected blackbody spectra\citep{dav05,dah06,dd08}.  We need excellent data in
order to detect these smeared spectral features, and use them to test
our understanding of the disk vertical structure.

The BHB LMC X-3 is the best object
currently known for this.  It has the lowest absorption column density
of any BHB which shows disk dominated spectra, extending the bandpass
over which the data can constrain the models down to the softest X-ray
energies. Additionally, the system parameters are fairly well
determined, especially distance which is known to better than 10 per
cent due to its location in the Large Magellanic Cloud. Thus the
conversion of flux to luminosity is subject to smaller uncertainties
than for any other BHB. This combination of properties means LMC X-3
offers a unique laboratory for testing our understanding of accretion
disk physics.

Here we use data from a Suzaku~\citep{mitsuda07} observation of the disk dominated state
in this object. This satellite combines moderate spectral resolution
CCD's useful to constrain 0.7-10 keV spectrum of the source, together with 
 the Si PIN photo-diodes of the non-imaging Hard 
X-ray Detector (HXD, Takahashi et al. 2007)
which can simultaneously determines the 12-30~keV
spectrum. Thus these data give a complete picture of the emission,
being able to constrain the (weak) tail to higher energies which
otherwise gives a significant source of uncertainty in reconstructing
the disk spectrum from CCD data alone.

The spectrum of LMC X-3 during this observation is dominated by a
clear disk component with luminosity around 10 per cent of the
Eddington limit. This is low enough to avoid the multiple
uncertainties which arise at higher luminosities, where the disk may
puff up due to radiation pressure, advection of radiation may become
important, and strong winds from the inner disk can distort the
spectrum both through absorption and from changing the intrinsic disk
spectrum due to the mass loss. Thus these observations give the most
sensitive test to date of our understanding of a `clean' accretion
disk to compare with the models. 

\section{Observation and Data reduction}

We observed LMC X-3 with Suzaku from 2008 December 22 UT07:14 through
December 23 UT20:49 (epoch 5).  Suzaku carries 4 X-ray telescopes
(XRT, Serlemitsos et al.~2007), each with a focal-plane X-ray CCD camera (X-ray Imaging
Spectrometer; XIS, Koyama et al.~2007) operating in the energy range
of 0.2--12 keV.  Three of the XIS (XIS0, 2, 3) have front-illuminated
(FI) CCDs, while XIS1 utilizes a back-illuminated (BI) CCD, achieving
an improved soft X-ray response but poorer hard X-ray
sensitivity. Since XIS2 is no longer available, we use the two
remaining FI-CCDs (XIS03) together with the BI-CCD (XIS1). We combine
these with data from the Si PIN photo-diodes of the HXD \citep{tt07}, 
which covers the 10--70 keV energy band.

The source was centered in the XIS field of view. The XIS were
operated with the 1/4 window option so that pileup did not become significant.  We use
version~2.2.11.22 of the pipeline provided by the Suzaku team, and the
XIS data are screened using the standard criteria (i.e. only GRADE0,
2, 3, 4, 6 events are accumulated, time interval after passage through
the South Atlantic Anomaly is larger than 436 seconds, the object is
at least $5^\circ$ and $20^\circ$ above the rim of the Earth during
night and day, respectively). This gives a net exposure of 73.98~ks.

We extract XIS events from both $5\times5$ and $3\times3$ editmodes using a circular
region with a radius of $4^\prime.3$ centered on the image peak. This
is larger than the window size so the effective extraction region is
the intersection of the rectangular window with this circle.  We use
{\sc ftool} {\sc xisrmfgen} ({\sc version 2007-05-14}) and {\sc xissimarfgen} ({\sc version 2009-01-08})to calculate the response
matrix and auxiliary response of the instrument. We co-add the data
from the two FI CCDs as these are very similar, and generate the
co-added response for this XIS03 spectrum using the {\sc ftools} {\sc
addrmf} and {\sc addarf}. We add systematic errors of 1\% to each
energy bin for both XIS03 and XIS1 spectra.

\begin{figure}
\epsscale{.80}
\plotone{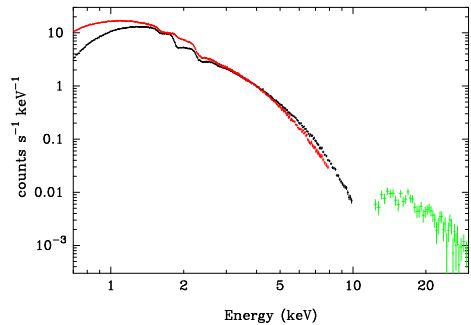}
\caption{The background subtracted 
X-ray spectra of LMC X-3 obtained with XIS03 (black), XIS1(red) and the HXD PIN(green). 
\label{fig:rawspec}}
\end{figure}

The source is bright, with count rate from XIS1 of 
$24~{\rm cts~s^{-1}}$ in the range of 0.7--10~keV, which corresponds to
$3.1\times 10^{-10}~{\rm erg~s^{-1}~cm^{-2}}$.
Though background can be
important towards the highest energy end of this bandpass,
the large point spread function of Suzaku means that the 1/4 window
data cannot be used to simultaneously determine the background as
there is no source-free region. Instead we use the Lockman hole data
obtained by Suzaku from 2009 June 12 UT 07:17 through June 14 UT01:31
with an exposure of a net exposure of 92~ks, using the same shaped
(intersection of circle and rectangle) region.  The background levels
are negligible in soft energy band, and reach to $\sim$0.7~\% and
$\sim5$~\% of our LMC X-3 data at 8~keV, for XIS03 and XIS1,
respectively.  In this paper, we used the 0.7--10~keV band and the
0.7--8~keV band for XIS03 and XIS1, respectively.  We show the
background subtracted XIS03 and XIS1 spectra in
Fig.~\ref{fig:rawspec}.

The HXD was operated in the nominal mode throughout the observations.
We use the standard pipeline processed PIN data (object is at least
$5^\circ$ above the earth rim, time interval after passage through the
South Atlantic Anomaly is longer than 500 seconds, and cutoff rigidity
is greater than 6~GV). We extract background from the same goodtime
intervals of the PIN background model based on {\sc lcfitdt} method ({\sc 2.0ver0804})
provided by the HXD team for each observation \citep{fukazawa09}. 
Here we note that the HXD-PIN background model is aimed to reproduce the
non X-ray background. 
The source exposure is corrected for the 7.3~\% deadtime, which is calculated from
the pseudo event rates \citep{tt07,kokubun07} by using the {\sc ftool}
{\sc hxddtcor}, giving a net exposure of 70.03~ks.  
This is $\sim 25\%$ less than the XIS,
but the source shows little variability so we do
not exclude any XIS data in order to have only exactly simultaneous
data. These data show that the source is not significantly detected at
the highest energies, so we restrict the fit range to 12--30~keV using
the appropriate response 
{\sc ae\_hxd\_pinxinome5\_20080716.rsp}.

The background subtracted PIN spectrum is shown in
Fig.~\ref{fig:rawspec} together with the XIS spectra. This has a
12-30~keV count rate of $0.076~{\rm cts~s^{-1}}$. 
This corresponds to
a flux of $2.2\times 10^{-11}~{\rm erg~s^{-1}~cm^{-2}}$, which is only
$\sim$ 4 times higher than flux of the cosmic X-ray background (CXB) in the
same energy range. Thus the CXB is not negligible, and we include this
as a fixed exponentially cutoff PL \citep{bolt87} in all our spectral
modeling for the PIN
data\footnote{http://heasarc.gsfc.nasa.gov/docs/suzaku/analysis/pin\_cxb.html}.

\section{Analyses and Results}

We describe here the system parameters of LMC X-3 adopted in this paper.  
LMC X-3 is a persistent source, so the emission from the outer
accretion disk always contributes to the optical spectrum, and X-ray
irradiation can alter the structure and spectrum of the companion
star. This introduces some uncertainty in determining the properties
of the companion, and hence the mass of the black hole \citep{klis86, cowley83, soria01,coe02}.
Taking account the uncertainties, we consider the black hole mass in the range of
7--9$M_\odot$ and inclinations of 50$^\circ$--67$^\circ$. Thus for each
physical model ({\sc kerrbb} and {\sc bhspec}\footnote{{http://www.sns.ias.edu/$\sim$swd/xspec.html}} in {\sc xspec}) we report fit values
over this range, showing also results for the specfic combination of
$7M_\odot$ and inclination of 67$^\circ$ used by \cite{ddb06}. 
We assume a source distance of 52~kpc \citep{di97}.

We use the {\sc xspec} spectral fitting package. 
Large fit residuals due to calibration uncertainties are often observed near the edge structures of the XIS/XRT instrumental responses.
We exclude energy range of 1.6--2.0~keV and 2.2--2.4~keV, 
and model the remaining instrumental feature at $\sim3.2$~keV  by
gold M edge by a gaussian line with fixed gaussian width of 0.1~keV as
is done in \cite{kubota07}.

The absorption column
density is well determined as $3.8\times
10^{20}$~cm$^{-2}$\citep{page03} from the depths of edges measured by the 
XMM-Newton Reflection Grating Spectrometer data. This is consistent
with most of the absorption being due to our own Galaxy as the
colum density along this line of sight is $3.2\times
10^{20}$~cm$^{-2}$ as measured by radio observations \citep{nowak01}.
However, low-energy calibration (especially the quantum efficiency) of the Suzaku XIS suffers from relatively large
systematic errors due to the uncertainties of the contamination thickness and its composition.   
To minimize the impact of these systematic errors while keeping good low-energy coverage, we restricted the energy range above 0.7~keV in the spectral analysis.   In this energy range
with the calibration data base ({\sc caldb: version xis20090203}), we estimate the systematic error in determining 
$N_{\rm H}$ is about a few times $10^{20}~{\rm  cm^{-2}}$ (XIS team, private communication).    Because this is comparable to the column density to LMC X-3, we allow $N_{\rm H}$ (described using {\sc phabs})
to be free in the spectral analysis unless otherwise stated.

When we analyze the disk emission, it is important to model the power-law component appropriately, because its low-energy part sometimes affect the fitting of the disk emission.
A simple power law (hereafter PL) plus CXB fit to the PIN data alone in the
12-30~keV band gives a photon spectral index of
1.66--2.38 with the best fit value of 2.01.
Because the power-law component in the high/soft state usually has a photon index larger than 2.0 (e.g., Done et al.~2007 and references therein), we consider that the true index of LMC X-3 lies between 2.00-2.38.     
Therefore, we constrain the index in this range in the spectral analysis rather than allowing it to be free.   Otherwise, the power-law component would be optimized to fill the fit residuals in the low-energy band, where the statistics are much better than those of the HXD PIN, and the best-fit power-law index could significantly deviate from the true slope in the hard energy band.

\begin{deluxetable}{ccrrrrrrrrc}
\tabletypesize{\scriptsize}
\tablecaption{The best fit parameters based on the {\sc diskbb} model\label{tab:diskb}}
\tablewidth{0pt}
\tablehead{
\colhead{model}  &   \colhead{$N_{\rm H}$}&\colhead{$T_{in}$}  
&\colhead{disk norm~$^a$}&
\colhead{$\Gamma~^b$} &\colhead{norm or $f$~$^c$} & \colhead{$\chi^2/dof$} \\
\colhead{} &    \colhead{$10^{20}~{\rm cm^{-2}}$}&\colhead{keV}  &\colhead{} &
& \colhead{} 
}
\startdata
diskbb + PL   &$6.04\pm0.02$ &$0.840\pm0.002$   &$26.0\pm0.3$ & $2.380^{+0}_{-0.007}$&$0.0354^{+0.0005}_{-0.0006}$  &$770.25/706$  \\
simpl$*$diskbb   &$ <7\times 10^{-3}$ &$0.787^{+0.003}_{-0.001} $     &$42.8^{+0.3}_{-0.5}$ & $2.380^{+0}_{-0.003} $& $0.115^{+0.0007}_{-0.0023}$ &$1336.14/706$  \\
\enddata
\tablecomments{Errors represent 90~\% confidence.}
\tablenotetext{a}{normalization is defined as 
$r_{\rm in}^2\cos i/(D/10~{\rm kpc})^2$.}
\tablenotetext{b}{Value of photon index $\Gamma$ is limited to be 2.0--2.38 (see text).}
\tablenotetext{c}{PL normalization at 1~keV in the unit of ${\rm photons~s^{-1}~keV^{-1}}$ or scattering fraction $f$ of the {\sc simpl} model.}
\end{deluxetable}

\begin{figure}
\epsscale{1}
\plotone{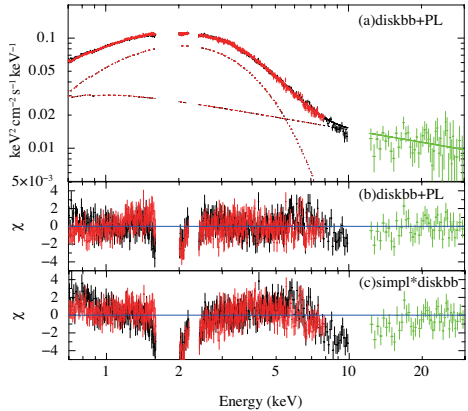}
\caption{$\nu F_\nu$ spectrum of LMC X-3 from the best fit {\sc diskbbb} plus {\sc PL} model  (a), 
residuals between the data and {\sc diskbb} plus {\sc PL} fit (b)
residuals between the data and best fit {\sc simpl}*{\sc diskbb} model.
In panel (a),  the CXB component is subtracted from the PIN spectrum. 
\label{fig:diskb}}
\end{figure}

\subsection{Empirical modeling of the disk: {\sc diskbb} } 

In order to compare the spectral data with the previous observations,
we first use the common black hole spectral model of an absorbed disk
plus power law. The simplest disk model is {\sc diskbb}, which has a
temperature distribution $T(r)\propto r^{-3/4}$, i.e. has no stress
free inner boundary condition \citep{mit84,max86}.

Figure~\ref{fig:diskb} shows the XIS and the HXD/PIN spectra with this
best fit {\sc diskbb} plus {\sc PL} model, with parameters given in
table~\ref{tab:diskb}. The spectrum is well fit
($\chi^2/dof=770.25/706$) by the dominant {\sc diskbb} component with
$T_{\rm in}$ of 0.84~keV.  The 0.7--30~keV flux is estimated as
$3.30\times 10^{-10}~{\rm erg~s^{-1}~cm^{-2}}$, which gives an
absorbed X-ray luminosity as $1.07\times 10^{38}~{\rm erg~s^{-1}}$ for
an isotropic emission at $D=52$~kpc.  This luminosity is about 10\% of
the Eddington luminosity for a $7~M_\odot$ black hole, and it is the
same level as that seen in the dimmer of the two ASCA observations
reported by \citep{kubota2005}.
The radius derived from the {\sc diskbb} normalization, 
$r_{\rm in}^2\cos i /(D/{\rm 10~kpc})^2= 26.0\pm0.3$ is also consistent with the
values of $25\pm3$ and $24\pm 2$ obtained from the ASCA observations
\citep{kubota2005}.

Even though the fit is good, a closer inspection of the figure shows
that this is not a good physical description of a model where the disk
provides the seed photons for Compton upscattering into the power law
tail as the PL extends below the disk at low energies. Hence we
replace the {\sc PL} model with the {\sc simpl} \citep{simpl} model
for Compton upscattering. This takes some
fraction of the disk seed photons and upscatters these to higher
energies. It is a convolution model, so requires an extended energy
range which we fix at 
0.1~keV to 1000~keV. 

The fit becomes substantially worse with this more physical model
($\chi^2/dof= 1336/706$), as shown by the residuals between the data
and the best fit {\sc simpl}*{\sc diskbb} model (Fig.~\ref{fig:diskb}c).  
In addition,  $N_{\rm H}$ is much smaller than expected, 
even though it is comparable to the current systematic uncertainty of
the $N_{\rm H}$ determination with the XIS.
Both these effects show that
the observed disk shape gives the excess at lower energies. 
In other words, a peak profile of the observed disk spectrum corresponding to the maximum disk temperature
is much broader than predicted by {\sc diskbb}, motivating us to do more detailed spectral analyses.  Similar
conclusions were reached by \cite{ddb06}, who analyzed BeppoSAX
data. Although their data had poorer signal-to-noise, it had similar
low energy coverage and they also found poor fits with {\sc diskbb}
once the PL was reasonably constrained.

\begin{deluxetable}{ccrrrrrrrrcrl}
\tabletypesize{\scriptsize}
\tablecaption{The best fit parameters based on the {\sc kerrbb}
model\label{tab:kerrbb}}
\tablewidth{0pt}
\tablehead{
\colhead{mass} & \colhead{$i$}&   \colhead{$N_{\rm H}$}&
  \colhead{$a^*$$^a$}
 &\colhead{$\dot{M}$~$^b$}&
\colhead{$\Gamma$} & \colhead{$f$ } & \colhead{$\chi^2/
dof$} \\
  \colhead{$M_\odot$} & \colhead{degree}&
\colhead{$10^{20}~{\rm cm^{-2}}$}&\colhead{}
& \colhead{$10^{18}~{\rm g/s}$} & \colhead{ } &
\colhead{} & \colhead{}

}
\startdata
7  &$67$&$ 0.16\pm0.14$   &$-0.07\pm0.01 $&$3.32\pm 0.03$ &
$2.38^{+0}_{-0.04} $ &  $  0.093^{+0.002}_{-0.004}$&
$783.58/706$  \\
%
%
7  &$50$&$ 0.8\pm0.1$   &$0.533^{+0.010}_{-0.007} $&$1.27\pm0.01$
&$2.37^{+0.01}_{-0.10} $ & $0.088^{+0.003}_{-0.008}$  &
$772.86/706$  \\
9  &$67$&$ 0.3^{+0.1}_{-0.1}$   &$0.29^{+0.01}_{-0.009}$&
$2.58\pm0.02$&$ 2.38^{+0}_{-0.06} $ &   $0.091^{+0.002}_{-0.005}$ &
$776.33/706$  \\
9  &$50$&$ 1.0\pm0.2$   &$0.738\pm 0.007$&$1.005\pm0.009$&$
2.31^{+0.07}_{-0.09} $ &   $0.082^{+0.008}_{-0.007}$ &
$776.70/706$  \\
\enddata
\tablecomments{Same as table~\ref{tab:diskb} but for the {\sc
kerrbb} model.}
\tablenotetext{a}{Blackhole spin parameter.}
\tablenotetext{b}{Mass accretion rate of the {\sc kerrbb} model.}
\end{deluxetable}

\subsection{Effect of the general relativity: temperature profile and relativistic smearing : {\sc diskpn}  and {\sc kerrbb}}

\begin{figure}
\plotone{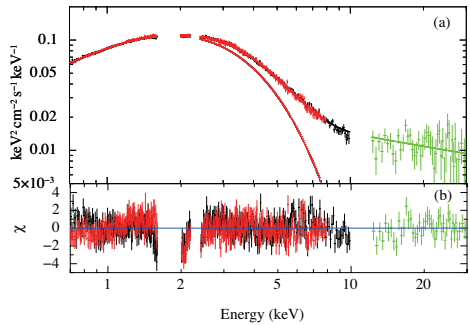}
\caption{$\nu F_\nu$ spectrum of LMC X-3 based on the best fit {\sc simpl}*{\sc kerrbb} model  (a), 
and residuals between the data and the model (b).
Same as Fig.~\ref{fig:diskb}, the CXB subtracted spectrum is shown for the PIN data (panel a). 
\label{fig:kerrbb}}
\end{figure}

We now use progressively more complex models for the disk emission
together with the {\sc simpl} Compton upscattering model for the tail.
First, we consider the {\sc diskpn} model which includes the
stress-free inner boundary condition assuming a Paczynski-Wiita
(pseudo-Newtonian) potential.  This smoothly connects the temperature
profile from $T(r)\propto r^{-3/4}$ at large radii, and 
$T(r_{\rm in})=0$ at $r_{\rm in}$, where $r_{\rm in}=6R_g$, in a way which is
similar to that expected from the fully relativistic equations for a
zero spin black hole \citep{gier99}. This has the effect of broadening
the spectrum slightly as it reduces the emission from the hottest
material, and hence gives a better fit to the data than the {\sc
simpl}*{\sc diskbb} ($\chi^2_\nu=984/706$), though still not as good
as the unphysical {\sc diskbb}+{\sc PL} model.  This indicates that
the soft component in the data is still broader than the {\sc diskpn}
description. Relativistic corrections to the temperature profile alone
do not sufficiently broaden the spectrum to match the observed data.

The next step up in complexity is to include the special and general
relativistic effects which distort the observed spectrum at each
radius. We use the {\sc kerrbb} model \citep{li05} for this, which
calculates the intrinsic temperature distribution from a fully
relativistic, stress-free inner boundary condition for arbitrary spin,
multiplied by a color temperature correction factor, $f_{\rm col}$, to
approximate the effect of modified blackbody emission. We fixed
$f_{\rm col}=1.7$ as appropriate for the XIS CCD bandpass at this
luminosity (Done \& Davis 2008). This intrinsic emission from each
radius is convolved with the relativistic transfer function for that
radius and assumed disk inclination\citep{li05}.

We explore the dependence on mass and inclination by fitting the model
with different combinations of these two parameters. These fit results
are shown in table~\ref{tab:kerrbb} and Fig.~\ref{fig:kerrbb}.  These
all give a fit which is similarly good as that derived from the {\sc
diskbb+PL} model (see table~\ref{tab:kerrbb}). The quality of the fit is not very
dependent on mass and inclination, but the derived value of the spin
changes from almost zero (lowest black hole mass, highest inclination)
to $\sim 0.7$ (highest black hole mass and lowest inclination).

We illustrate the fit in Fig.~\ref{fig:kerrbb}a, showing the $\nu
F_\nu$ spectrum and residuals for the models with mass of $7M_\odot$
and inclination of $67^\circ$, as used by \cite{ddb06}. 
This gives a good fit ($\chi^2_\nu=776/706$) with flat residuals
across the whole energy band (see Fig.~\ref{fig:kerrbb}b), 
and gives a spin of $a^*\simeq 0$, consistent with that derived by \cite{ddb06}.
Relativistic effects on the disk continuum were first proposed by \cite{cunningham75}, but this is the first time they 
have been significantly detected, as it is much harder to disentangle
this smearing on a broad continuum than on a line \citep{fabian89}.
Though the smaller values of $N_{\rm H}$ may suggest that the data still has excess soft emission and is somewhat broader even than that produced by the temperature profile and relativistic broadening, 
the calibration uncertainties do not enable us to conclude unambiguously 
whether there are other broadening factors.

\subsection{ Effect of the vertical structure of the disk : {\sc bhspec}}

The assumption of a constant $f_{\rm col}$ is only an approximation to the true 
spectrum from a given radius.  In the real disk,  changes in opacity at 
atomic edges produce spectra intrinsically broader than a blackbody.  Under 
the assumption of the solar metallicity, the {\sc bhspec} model \citep{dav05} 
uses radiative transfer to calculate the intrinsic spectrum at a given radius, 
so does self-consistently includes the color temperature correction. It then 
convolves this with the relativistic transfer functions to produce the best disk 
spectra to date.

\begin{figure}
\plotone{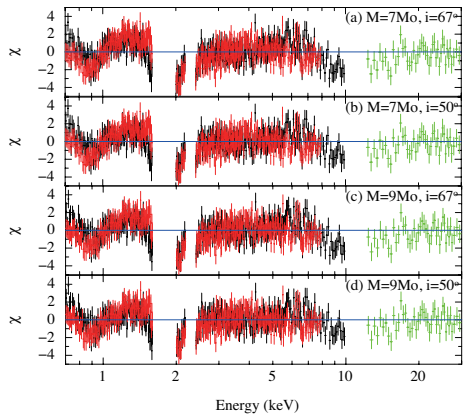}
\caption{Residuals between the data and the best fit {\sc bhspec} models with different mass and inclination. 
(a)$M=7M_\odot$ and $i=67^\circ$,  
(b)$M=7M_\odot$ and $i=50^\circ$,  
(c)$M=9M_\odot$ and $i=67^\circ$,  and
(d)$M=9M_\odot$ and $i=50^\circ$. 
\label{fig:bhspec}}
\end{figure}

We fit the spectra with the {\sc bhspec} disk model for the same
combinations of $i$ and $M$ as for the {\sc kerrbb} fits above. The
results are shown in table~\ref{tab:bhspec}.
The model gives similar spins to
those derived from {\sc kerrbb} (see table~\ref{tab:kerrbb}), 
showing that the assumption of $f_{\rm col}=1.7$ is 
appropriate to the {\sc kerrbb} fit for the XIS CCD bandpass.
Moreover, the estimated absorption 
is now consistent with the known column density in this direction $\sim 3\times 10^{20}~{\rm cm^{-2}}$.
Thus the {\sc bhspec} model produces additional excess soft emission 
than produced by relativistic broadening alone.
However, despite being a more physical model, it gives a 
much {\em worse} fits than {\sc kerrbb}
($\chi^2/dof=1212/706$ for the case of $M=7M_\odot$ and
$i=67^\circ$ compared to $783/706$).  Figure~\ref{fig:bhspec} shows the
residuals between the data and the model for each mass and inclination. 
These are all clearly
dominated by a broad feature below 1~keV,
which is caused by the smeared atomic absorption edges present in the model.



We  checked the result by including relativistically smeared ionised
reflection, modelled using the convolution version of the models of
\cite{bal01} as described in \cite{done06}.
Though the reflection itself cannot be well constrained because of couplings to 
mass, inclination and the difference of the continuum disk model, 
it does not change this
conclusion: {\sc bhspec} gave a worse fit to the data than {\sc kerrbb}
if we did not constrain its absorption, 
and the {\sc bhspec} residuals are dominated by features around 1~keV.

\begin{deluxetable}{ccrrrrrrrrcrl}
\tabletypesize{\scriptsize}
\tablecaption{The best fit parameters based on the {\sc bhspec} model\label{tab:bhspec}}
\tablewidth{0pt}
\tablehead{
 \colhead{mass} & \colhead{$i$}&   \colhead{$N_{\rm H}$}& 
 \colhead{$a^*$} &\colhead{$L_{\rm disk}$} &
\colhead{ $\Gamma$} &\colhead{$f$}  & \colhead{$\chi^2/dof$} \\
 \colhead{$M_\odot$} & \colhead{degree}&   \colhead{$10^{20}~{\rm cm^{-2}}$}&\colhead{} 
&  \colhead{$10^{38}~{\rm erg/s}$ } &&
\colhead{} 

}
\startdata
\multicolumn{8}{c}{the solar abundance}\\
\hline
7  &$67$ &$ 2.7\pm0.1$  &$-0.082^{+0.012}_{-0.008} $ &
$1.799\pm0.005$
 &$ 2.380^{+0}_{-0.006} $&   $ 0.102^{+0.001}_{-0.002}$&$1212.47/706$  \\
7  &$50$ &$ 3.9\pm0.1$  &$0.542\pm0.006 $ 
&$1.054^{+0.002}_{-0.003}$   &$ 2.380^{+0}_{-0.008} $&  $  0.096\pm0.002$&$1058.56/706$  \\
9  &$67$ &$ 3.1\pm0.1$  &$0.286 \pm0.009 $ &
$1.748 \pm 0.005$
 &$ 2.380^{+0}_{-0.006} $&   $ 0.100\pm0.001$& $1094.30/706$  \\
%
%
9  &$50$ &$ 4.3\pm0.1$  &$0.742 \pm0.004$ &
$1.085^{ +0.002}_{-0.003}$
 &$2.380^{+0}_{-0.010} $&   $ 0.094\pm0.002$& $974.65/706$  \\
%
%
\hline
\multicolumn{8}{c}{1/3 solar abundance}\\
\hline
7  &$67$ &$ 1.6 \pm0.1$  &$-0.076\pm0.010 $ &
$1.765^{+0.005}_{-0.004}$
 &$ 2.380^{+0}_{-0.007} $&   $ 0.102\pm0.002$&$1151.91/706$  \\
7  &$50$ &$ 2.7 \pm0.1$  &$0.536\pm0.006 $ &
$1.033\pm0.002$
 &$ 2.380^{+0}_{-0.008} $&   $ 0.098\pm0.002$&$1062.58/706$  \\
9  &$67$ &$ 2.0 \pm0.1$  &$0.278\pm0.009 $ &
$1.717\pm0.005$
 &$ 2.380^{+0}_{-0.006} $&   $ 0.102\pm0.002$&$1063.26/706$  \\
9  &$50$ &$ 3.1 \pm0.1$  &$0.737\pm0.004 $ &
$1.061\pm0.002$
 &$ 2.380^{+0}_{-0.009} $&   $ 0.096\pm0.002$&$987.75/706$  \\
\hline
\multicolumn{8}{c}{1/10 solar abundance (just for comparison)}\\
\hline
7  &$67$ &$ 0.7 \pm0.1$  &$-0.049\pm0.010 $ &
$1.725^{+0.005}_{-0.004}$
 &$ 2.380^{+0}_{-0.008} $&   $ 0.102\pm0.002$&$1014.46/706$  \\
7  &$50$ &$ 1.69 \pm0.1$  &$0.552^{+0.007}_{-0.006} $ &
$1.011^{+0.002}_{-0.003}$
 &$ 2.380^{+0}_{-0.008} $&   $ 0.099\pm0.002$&$974.17/706$  \\
9  &$67$ &$ 1.0^{+0.2}_{-0.1}$  &$0.304^{+0.009}_{-0.010} $ &
$1.680\pm0.005$
 &$ 2.380^{+0}_{-0.009} $&   $ 0.101\pm0.002$&$973.39/706$  \\
9  &$50$ &$ 2.0 \pm0.1$  &$0.748\pm0.004 $ &
$1.039^{+0.002}_{-0.003}$
 &$ 2.380^{+0}_{-0.010} $&   $ 0.097\pm0.002$&$919.16/706$  \\
\enddata
\tablecomments{Same as table~\ref{tab:diskb} and table~\ref{tab:kerrbb} but for the {\sc bhspec} model.}
\end{deluxetable}

\subsection{ Constraint on the absorption lines due to highly ionized iron}

Absorption lines from highly ionized Hydrogen (H-) and Helium(He-)
like iron are often seen in disk dominated black hole binaries (see
e.g. the compilation in \citealt{done07}). We
focus on the iron K line region to sensitively search for such
features. We fit the 5-8~keV data with a power law and narrow
absorption line (with width fixed at 10~eV). There is evidence for
such a line at $6.61\pm 0.04$~keV, of equivalent width of $-15\pm 6$~eV. 

Though this is significant at about the 99\% confidence level on an
F-test as $\chi^2_\nu$ drops from $233/221$ to $215/219$, the
line energy is slightly too low to be He-like iron at 6.70~keV, 
even if we consider the calibration uncertainty of the gain, $\sim 20$~eV, 
when the 1/4 window option is used\footnote{Based on the {\sc caldb} version {\sc xis20090203}}.   
 The difference may be even larger, because the lines are often blue shifted due to the disk wind.
As seen in the Chandra/HETGS spectra of GX~13+1~\citep{ueda13+1} and GRS~1915+105~\citep{ueda1915}, 
He-like iron energy can be observed at lower energies than the prediction. 
\cite{ueda1915} suggested that the slight shift of the line
energy can be caused by a possible contamination of less ionized ions
such as Li-like ones at 6.68 keV, uncetainties in the incident line
energies for He-like (and more electrons) iron, and/or the
velocity-field structure in the wind along the line-of-sight.
However, even considering these effects and the XIS calibration uncertainty, 
$6.61\pm0.04$~keV is still difficult to be interpreted as He-like iron absorption lines. 
Fixing the line energy at 6.70~keV gives $\chi^2_\nu=226/220$,
so again formally significant at the $\sim$~99\% level, with an
equivalent width of $-9\pm 6$~eV.  
This value represents the conservative constraint on the He-like iron absorption line. 

Other BHB at similar continuum luminosities ($\sim 10^{38}~{\rm
erg~s^{-1}}$) show absorption lines of equivalent width of $\sim30$~eV
in both H- and He-like iron (e.g., GRO~J$1655-40$
\citep{ueda98,yamaoka01,miller06}, 1H~$1743-322$\citep{h1743}).
These BHBs all have similar
inclination angle of $i\sim 70^\circ$.  Thus less
significant absorption lines may indicate that the inclination is
likely to be lower than $67^\circ$.  Alternatively, these BHBs showed
higher disk temperature $T_{\rm in}$ than our LMC X-3 data, typically
at $\ge 1.0$~keV when they showed absorption lines, and thus the disk
temperature may also be important in creating the disk wind in
addition to the luminosity.

\begin{figure}
\plotone{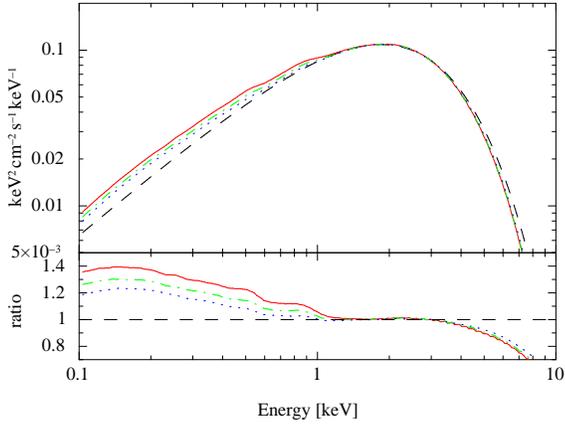}
\caption{Comparison of the best fit {\sc bhspec} and {\sc kerrbb} (black dashed line) with 
$i=67^\circ$ and $M=7~M_\odot$. 
The bhspec models are shown with 1-solar abundance (red solid line), 1/3-solar abundance (green dot-dashed line) and 
1/10-solar abundance (blue dot line). 
Ratios of the best fit {\sc bhspec} models to the {\sc kerrbb} model are shown in the bottom panel.
For the plotting purpose, the interstellar absorption and {\sc simpl} are excluded in both 
models.
 \label{fig:model}}
\end{figure}
\section{Discussion \& Conclusions}

\subsection{Black hole spin}

The resulting values of spin are fairly similar for both {\sc bhspec}
and {\sc kerrbb}  with $f_{\rm col}=1.7$, giving a range 
in spin from 0--0.7 depending on mass
and inclination. Although this seems a large range, it only
corresponds to a change in inner radius by less than a factor of 2,
from $6R_g$ at $a^*=0$ to $3.4R_g$ at $a^*=0.7$, whereas for extreme
spin at $a^*=0.998$ the inner radius becomes as small as $1.23R_g$. 
Thus the models are not very sensitive to spin for low to moderate spin, even for the
best constrained black hole system parameters, as the implied change
in radius for the last stable orbit is low. However, they are very
sensitive to high spin, and strongly rule out such values for LMC X-3.

Hence we conclude that the black hole in LMC X-3 has low--to--moderate
spin. This is also the case for all other spin determinations from disk 
continuum fits \citep{ddb06,shafee06,gou09},
apart perhaps from GRS~$1915+105$(McClintock et al 2006 but see Middleton et al 2006). 
Such low to moderate spins are in line 
with current (but probably quite uncertain) theoretical models which indicate birth
spins of $\la 0.8-0.9$ from stellar collapse \citep{gamie04}. However, we note that
the iron line fits often imply rather higher spins, sometime for the same objects
(e.g. Miller et al 2009). This is clearly an area of current intense research, 
see e.g. the discussion in \cite{kol09}.

\subsection{Vertical Structure}

The Suzaku observation of LMC X-3 shows a spectrum which is dominated
by the accretion disk, giving the best data yet on its detailed
spectrum,  and a simultaneous determination of the high energy tail.
We model the tail as
Compton upscattering of seed photons from the disk.  The disk spectrum
is much broader than a simple sum of blackbodies with $T(r)\propto r^{-3/4}$ ({\sc diskbb}). It is also broader than a sum of blackbodies
with temperature profile given by the (approximate) stress-free inner
boundary condition in the pseudo-Newtonian potential ({\sc
diskpn}). Relativistic smearing of the intrinsic blackbody spectra
from each radius (with temperature profile given by the stress-free
inner boundary condition) is required  before the model gives a good
description of the overall shape of the data ({\sc kerrbb}),
though the {\sc kerrbb} model also requires an absorption column close to zero.


The {\sc bhspec} model dispenses with the color temperature correction by
calculating the spectrum self-consistently from radiative transfer
through the vertical structure of the disk. The column density derived
from this model is consistent with the known column, but
the fit is poor due to residuals at 1~keV.  In order to clarify the
reason for the residuals seen in the {\sc bhspec} fit,
Fig.~\ref{fig:model} shows a comparison of the best fit {\sc bhspec}
model with that of {\sc kerrbb}.  The {\sc bhspec} model 
gives more soft emission below 1~keV,
and it does this due to the presence of absorption edges, 
most importantly the K edges of H-like
Nitrogen and Oxygen at 0.66 keV and 0.87 keV, respectively.  
Though these are broadened and smeared by the relativistic effects,
they are still deeper than any features present in the data, so appear in the
residuals as an excess at the absorption edge energy. These residuals
are around 5\% (peak to peak), but the data are actually consistent
with no structure at these energies, as seen by the featureless
residuals from the best fit {\sc kerrbb} model (Fig.~\ref{fig:kerrbb}b).

\begin{figure}
\plotone{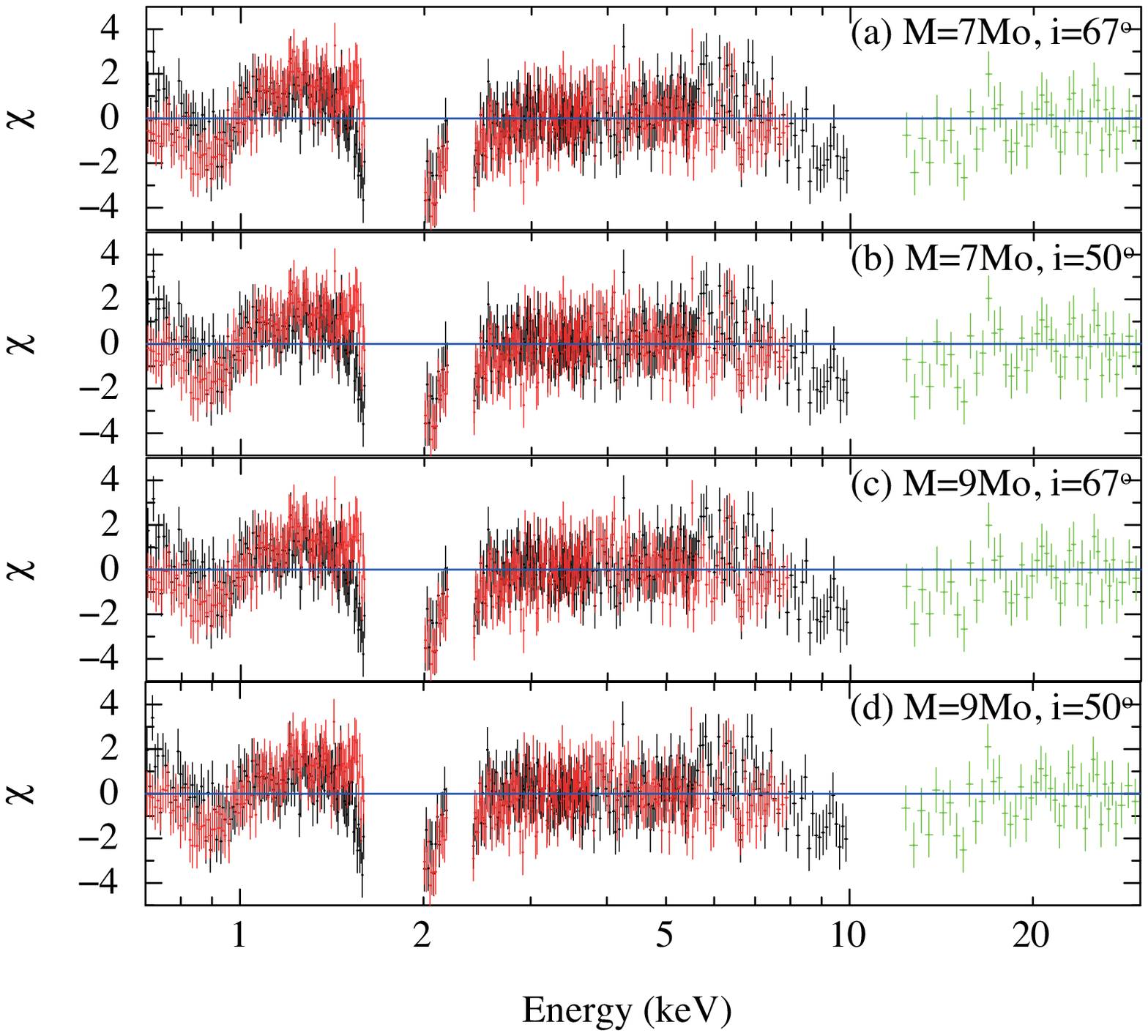}
\caption{ Same as Fig.~\ref{fig:bhspec} but for 1/3-solar abundance.
 \label{fig:033metal}}

\vspace{8mm}

\plotone{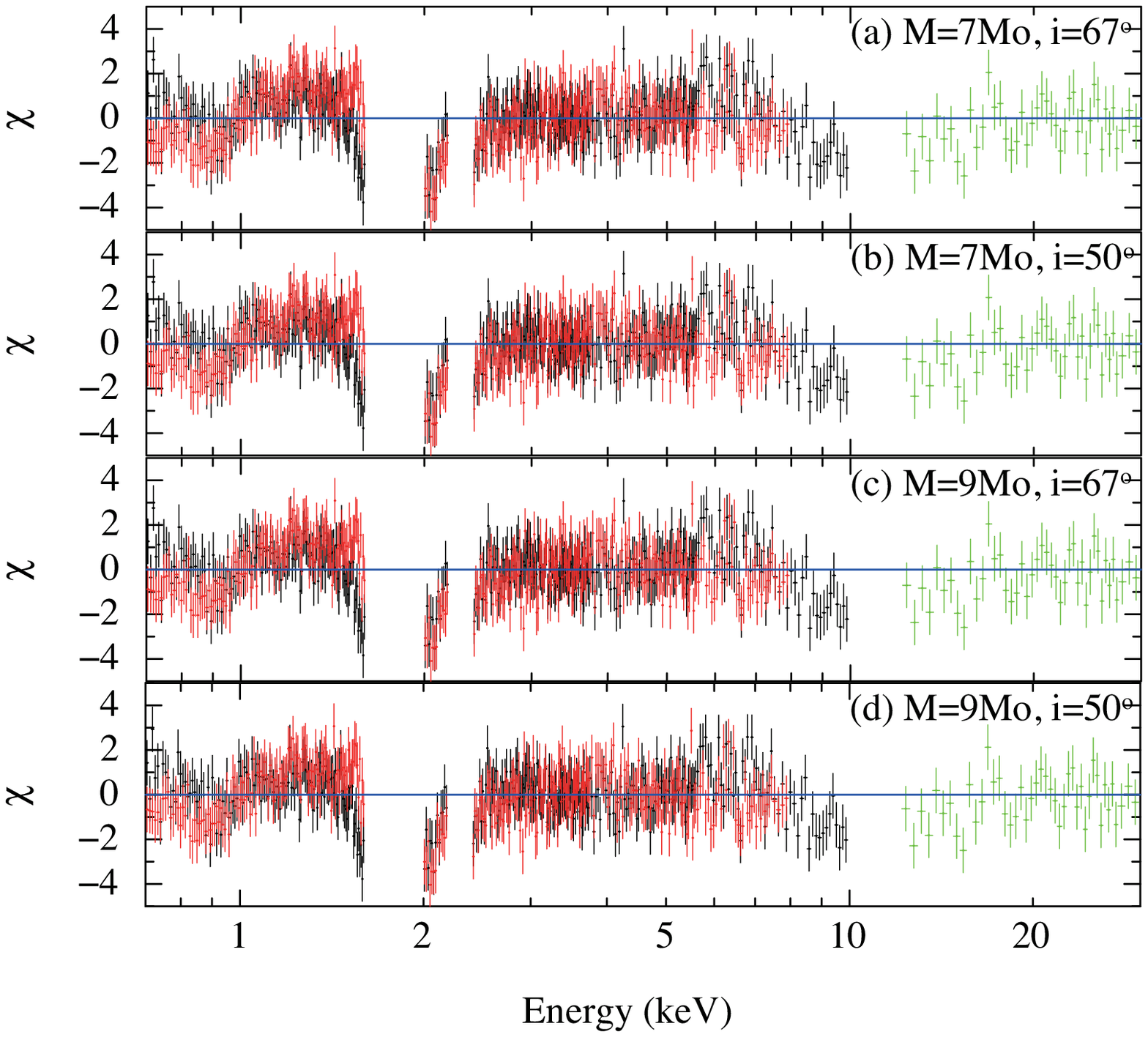}
\caption{ Same as Fig.~\ref{fig:bhspec} but for 1/10-solar abundance.
 \label{fig:01metal}}
\end{figure}

To reduce the size of the atomic features, 
we first consider sub-solar abundances, since 
the LMC is suggested to have $1/2\sim 1/3$ solar abundance~\citep{russell89,russell90,korn02}.
In order to examine this effect, we extended the original {\sc bhspec} model 
to account for sub-solar metallicity.  
We fit the data with the new {\sc bhspec} disk model of 
fixed 1/3-solar metallicity for the same combinations of $i$ and $M$ as used previously. 
The results and fit residuals are shown in table~\ref{tab:bhspec} and Fig.~\ref{fig:033metal}, 
respectively. For the condition of $M=7~M_\odot$ and $i=67^\circ$, 
the best fit {\sc bhspec} model with 1/3-solar abundance is 
shown in Fig.~\ref{fig:model} together with the original {\sc bhspec} and {\sc kerrbb} models.
The lower abundance means that the absorption 
edge structure below 1~keV is less significant 
($\sim3$\% peak-to-peak), which makes the fit slightly better ($\chi^2_\nu=1152/706$
for $M=7~M_\odot$ and $i=67^\circ$), 
though the residuals are still clearly 
dominated by the edge structure below 1~keV.
We also show the fit results based on the {\sc bhspec} with 1/10-solar abundance
in table~\ref{tab:bhspec}, Fig.~\ref{fig:01metal}, and Fig.~\ref{fig:model}.
Even though this is an underestimate of the metallicity, 
the predicted edge features are still larger than those seen in the 
data (Fig.~\ref{fig:01metal}), and thus the fits are worse than the {\sc kerrbb}. 

This illustrates the key issue. The data indicate that the spectrum is broader than
a single color temperature corrected, relativistically smeared model
({\sc kerrbb}). Including ion opacities as in {\sc bhspec} gives broader spectra, but
with the inevitable consequence edge features are predicted. 
Since there is no evidence for these edges in the spectrum, incorporating
sub-solar abundances alone in {\sc bhspec} 
cannot explain why the model does not reproduce the data.

Instead, this could point to the importance of including more atomic physics 
in {\sc bhspec}. Currently it only uses bound-free opacity, but 
bound-bound (line) transitions could also be important,
especially if there is large scale turbulence/convective motion in the disk atmosphere.  
However, if these additional transitions primarily show up as absorption features, they 
will further enhance the depth of the atomic absorption, increasing the discrepancies.  
Instead, line and/or radiative recombination continuum emission could fill in 
some of the absorption features, while still retaining the broader continuum, though 
this seems somewhat fine-tuned. Self-irradiation of the disk may be a better 
mechanism,  as this would drive the photosphere towards
isothermality, removing the atomic features which do not appear to be
present in the data but keeping the broader continuum.

Whatever the origin of the fit residuals, their existence demonstrates
the possibility of using the observational data to discriminate between
different models of the vertical structure of the accretion disk, and
motivates development of more sophisticated disk atmosphere
calculations.

\vspace{8mm}
We are grateful to all the Suzaku team members. 
The present work is supported by Grant-in-Aid No.19740113 from Ministry of 
Education, Culture, Sports, Science and Technology of Japan. 
SWD acknowledges support from the IAS, through grants NSF AST-0807432, NASA NNX08AH24G, and NSF AST-0807444.







\clearpage

\end{document}